\documentclass[aps,prb,reprint,groupedaddress,showpacs,amsmath,amssymb,twocolumn,floatfix]{revtex4}
\usepackage{txfonts}
\usepackage{graphicx,amssymb,mathrsfs,bm,color,times}
\usepackage{ulem}

\newcommand{\be}{\begin{equation}}

\makeatletter

\newcommand{\Rmnum}[1]{\expandafter\@slowromancap\romannumeral #1@}
\makeatother

\begin{document}

\title {Synthesizing Lattice Structure in Phase Space  }

\author{Lingzhen Guo$^{1,2}$}
\author{Michael Marthaler$^{1,2}$}
\affiliation{$^1$\mbox{Institut f\"ur Theoretische Festk\"orperphysik, Karlsruhe Institute of Technology, 76128 Karlsruhe, Germany}\\
$^2$\mbox{ DFG-Center for Functional Nanostructures (CFN),
Karlsruhe Institute of Technology, 76128 Karlsruhe, Germany}\\}

\date{\today}

\begin{abstract}
We consider a realistic model, i.e., ultracold atoms in a driven
optical lattice, to realize phase space crystals \cite{Guo}. The
corresponding lattice structure in phase space is more complex and
contains rich physics. A phase space lattice differs fundamentally
from a lattice in real space, because its coordinate system, i.e.,
phase space, has a noncommutative geometry, which naturally
provides an artificial gauge (magnetic) field. We study the
behavior of the quasienergy band structure as function of the
artificial magnetic field and investigate the thermal properties.
Synthesizing lattice structures in phase space is not only a new
way to create artificial lattice in experiments but also provides
a platform to study the intriguing phenomena of driven systems far
away from equilibrium.
\end{abstract}

\pacs{67.85.-d, 42.65.Pc, 03.65.-w, 05.45.-a}

\maketitle

\section{Introduction}

In a recent paper \cite{Guo}, we introduced the idea of phase
space crystals, i.e., a lattice structure in phase space created
by breaking a continuous phase rotational symmetry  via a driving
field. In our previous work we used the model of ultracold atoms
trapped in a time-dependent power-law potential, i.e., $\sim
x^n\cos(\omega_dt)$, to illustrate our idea. However, this model
is technically difficult to realize in experiments. Here, we
present a realistic driven optical lattice model, i.e, the
power-law driving is replaced by a cosine-type driving, i.e.,
$\sim \cos(kx+\omega_dt)$, to realize phase space crystals. Thus,
the novel phenomena predicted by phase space crystals can be
directly observed in current experiments of ultracold atoms in an
optical lattice.

The model proposed here synthesizes a more complex lattice
structure in phase space and thus contains rich physics. We
further develop the theory of phase space crystals and calculate
the complex quantum tunnelling rates. We identify  the artificial
(magnetic) gauge  field in phase space, which is a result of the
noncommutative geometry of the phase space crystal. Compared to
the artificial lattice structures in real space
\cite{photonic1,photonic2,photonic3,photonic4,phononic,Metamaterials1,Metamaterials2,Metamaterials3},
synthesizing a lattice structure in phase space has the key
advantage of being
 conveniently tunable in experiments through changes in the driving field. Due
to this possibility phase space lattices may provide a new
platform to simulate condensed matter phenomena.

\begin{figure}
\centerline{\includegraphics[scale=0.7]{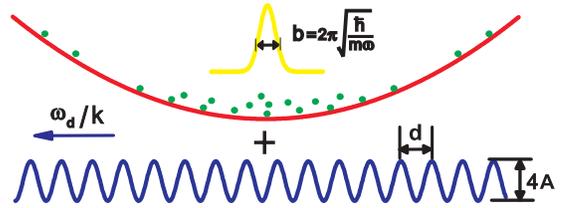}}
\caption{\label{fig DrivenOpticalLattice}{\bf{ Ultracold atoms in
driven optical lattice.}} Ultracold atoms (green dots) are
confined in a harmonic potential (red parabolic curve). The ground
sate of confinement potential is represented by a Gaussian wave
packet (yellow wave packet) with width $b=\sqrt{\hbar/(m\omega)}$.
The blue curve represents a propagating optical lattice with
period $d$, amplitude $2A$ and velocity $\omega_d/k$. The
potential for creating phase space lattice is the sum of them. }
\end{figure}

\section{Model and Hamiltonian}

The model we propose here can be realized by ultracold atoms
trapped in a time-dependent optical lattice. The Hamiltonian is
given by
\begin{equation}\label{DrivenOpticalLattice}
H(t)=\frac{p^2}{2m}+\frac{1}{2}m\omega^2x^2+2A\cos(kx+\omega_dt).
\end{equation}
Here, the  parabolic term is the harmonic confinement
potential of ultracold atoms, which can be created by a gaussian beam
profile of a laser \cite{Bloch2005Nature}  or introduced by
another external field. As sketched in Fig.~\ref{fig
DrivenOpticalLattice}, the characteristic length of the ground
state in the confinement potential is
$b=2\pi\sqrt{\hbar/(m\omega)}$. Experimentally  the optical
lattice is created by the interference of two counter-propagating
laser beams, which form an optical standing wave with period
$d=2\pi/k$. The ultracold atoms are trapped by the interaction
between the laser light field and the oscillating dipole moment of
atoms induced by the laser light \cite{Bloch2008RevModPhy}. We can
drive the optical lattice simply by tuning the phase difference of
the two laser beams linearly as described by Hamiltonian
(\ref{DrivenOpticalLattice}). Effectively, this creates a
propagating optical lattice with a velocity of $\omega_d/k$.
An important parameter is $\lambda\equiv (b/d)^2=\hbar
k^2/(m\omega)$, which defines the ``quantumness" of our system. It
is large in the quantum regime and goes to zero in the
semiclassical limit. We emphasize that the optical potential is
time-dependent and the confinement potential also plays an
important role. Thus, our system does not have spatial periodicity and
the Bloch theory in real space does not apply directly for the
Hamiltonian (\ref{DrivenOpticalLattice}).

\begin{figure*}[tp]
\begin{center}
\begin{tabular}{lll}
    \textbf{(a)} & \textbf{(b)}& \textbf{(c)}\\
    \includegraphics[scale=1.0]{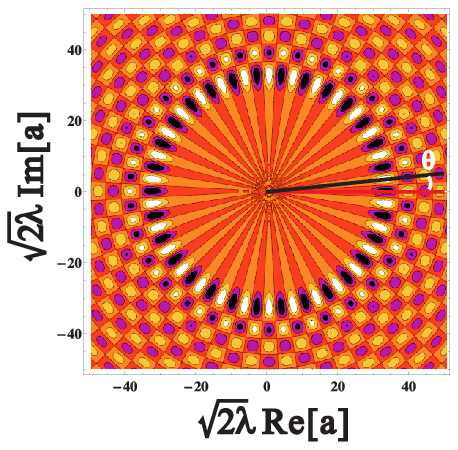} &
    \includegraphics[scale=1.0]{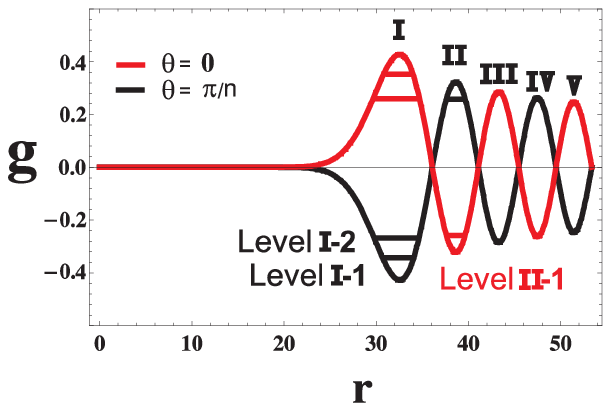} &
    \includegraphics[scale=1.0]{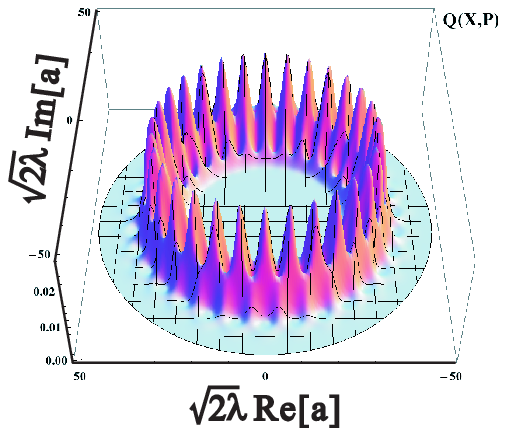}
\end{tabular}
\caption{\footnotesize{\textbf{Phase space lattice with
$\textbf{n=30}$}:  a) Lattice structure with $n$-fold symmetry in
phase space by plotting Hamiltonian (\ref{HRWA}) in the
semiclassical limit. The color represents the value of quasienergy
$g$, which indicates the whole lattice is composed of two
identical sublattices because of the chiral symmetry for
$\delta\omega=0$. b) Quasienergy along the radius direction with
angle $\theta=0$ and $\theta=\pi/n$. The roots of $g(r)=0$ divide
the whole lattice into many loops, which are labelled by Roman
numerals as indicated on the top of this figure. The corresponding
localized quantum states inside the loops are labelled by $Level\
\Rmnum{1}$-1, $Level\ \Rmnum{1}$-2, $Level\ \Rmnum{2}$-1, etc. c)
$Q$-function of a quasinumber state with quasinumber $m=0$. The
periodicity of $Q$-function reflects the $n$-fold symmetry of
phase space lattice.}} \label{fig PhaseSpaceLatticen30}
\end{center}
\end{figure*}

We are interested in the regime near the high-order resonant
condition $\omega_d\approx n\omega$ with a large integer $n\gg1$.
For the duration of this paper with will use $n=30$. The detuning
$\delta\omega\equiv \omega -\omega_d/n$ is much smaller than the
natural frequency $\omega$. We perform a unitary transformation of
the Hamiltonian $H(t)$ via the operator
$\hat{U}=e^{i(\omega_d/n)\hat{a}^\dagger \hat{a}t}$, where
$\hat{a}$ is the annihilation operator of the oscillator. In the
spirit of the rotating wave approximation (RWA), we drop the fast
oscillating terms and arrive at the time-independent Hamiltonian
(see more details in section A of the Appendix)
\begin{eqnarray}\label{HRWA}
\hat{g}&=&\lambda\epsilon (\hat{a}^\dagger
\hat{a}+\frac{1}{2})+\mu
\Big[\Big(\frac{\lambda}{2}\Big)^{-\frac{n}{2}}e^{-\frac{\lambda}{4}-i\frac{n\pi}{2}}\hat{a}^n\
L_{\hat{a}^\dagger
\hat{a}}^{(-n)}\Big(\frac{\lambda}{2}\Big)+h.c.\Big].\nonumber\\
\end{eqnarray}
In the context of Floquet theory, $\hat{g}$ is called
\textit{quasienergy} \cite{Haenggi,ME1Dykman}, which has been scaled by the energy
$m(\omega/k)^2=\hbar\omega/\lambda$. The parameters
$\epsilon\equiv\delta\omega/\omega$ and $\mu\equiv\lambda
A/(\hbar\omega)$ are the dimensionless detuning and driving
strength respectively. Functions $L_{\hat{a}^\dagger
\hat{a}}^{(-n)}(\bullet)$ are the generalized Laguerre
polynomials, as a function of the photon number $\hat{a}^\dagger
\hat{a}|k\rangle = k |k\rangle$, where $|k\rangle$ are the Fock
states.

\section{Symmetries}

In the following, we are particularly interested in the resonant
condition, i.e., the detuning is zero $\delta\omega=0$. Without
loss of generality, we set the scaled driving strength to unity,
i.e., $\mu=1$. In this case, the RWA Hamiltonian (\ref{HRWA}) has
two new symmetries which are not visible in the original
Hamiltonian (\ref{DrivenOpticalLattice}). To visualize them, we
replace the operator $\hat{a}$ by a complex number in the
semiclassical limit and plot the quasienergy $g$ in the phase
space spanned by Re$[a]$ and Im$[a]$. As displayed in
Fig.~\ref{fig PhaseSpaceLatticen30}(a), we first see the
\textit{discrete angular symmetry} $g(\theta)=g(\theta+2\pi/n)$.
Additionally we have the \textit{chiral symmetry}
$g(\theta)=-g(\theta+\pi/n)$, which divides the whole lattice
structure into two identical sublattices as indicated in
Fig.~\ref{fig PhaseSpaceLatticen30}(a) by the different colors. To
describe the two symmetries in quantum mechanics, we define a
unitary operator $\hat{T}_\tau=e^{-i\tau {\hat{a}^\dagger}
\hat{a}} $ with the properties $\hat{T}_\tau^\dagger \hat{a}
\hat{T}_\tau = \hat{a} e^{-i\tau}$ and $\hat{T}_\tau^\dagger
\hat{a}^n \hat{T}_\tau = \hat{a} e^{-in\tau}$. Since the operator
$\hat{a}^\dagger\hat{a}$ keeps invariant under the transformation
of $\hat{T}_\tau$, it is not difficult to check that the RWA
Hamiltonian (\ref{HRWA}) is invariant under discrete
transformation $T_\tau^\dagger \hat{g} T_\tau = \hat{g}$ for
$\tau={2\pi}/{n}$. We call this symmetry \textit{discrete phase
translation symmetry}. The chiral symmetry follows from the fact
$T_\tau^\dagger \hat{g} T_\tau = -\hat{g}$ for $\tau={\pi}/{n}$.
The chiral symmetry suggests that the two sublattices are
symmetric with respect to $g=0$, except a phase shift
$\theta\rightarrow\theta+\pi/n$. The angular symmetry indicates it
is convenient to introduce the radial and angular operators
$\hat{r}$ and $\hat{\theta}$ via
$\hat{a}=e^{-i\hat{\theta}}\hat{r}/\sqrt{2\lambda}$ and
$\hat{a}^\dagger=\hat{r}e^{i\hat{\theta}}/\sqrt{2\lambda}$. They
obey the commutation relation
\begin{equation}\label{comm}
[\hat{r}^2,e^{i\hat{\theta}}]=2\lambda e^{i\hat{\theta}}
\end{equation}
where $\lambda$ plays the role of a dimensionless Plank constant.

\section{Phase space lattice}

In the semiclassical limit $\lambda\rightarrow 0$, the quantum
Hamiltonian $\hat{g}$ can be written in its classical form (see
more details in section A of the Appendix)
\begin{equation}\label{g}
g=\frac{1}{2}\epsilon r^2+2\mu J_n(r)\cos(n\theta-\frac{n\pi}{2}).
\end{equation}
Here, we have used the asymptotic property of Laguerre
polynomials, i.e., $\lim_{k\rightarrow \infty}L_k^{(n)}(x/k)=k^n
e^{\frac{x}{2k}}x^{-n/2}J_n(2\sqrt{x})$, where $J_n(\bullet)$ is
the Bessel function of order $n$. The angular periodicity comes
from the cosine function in Eq.(\ref{g}) while the radial
structure is created by the Bessel function $J_n(r)$. A similar
situation has recently been studied in voltage biased Josephson
junctions \cite{AnkerholdJJOsci,ArmourJJOsci}. The zero lines of
$g$ form the `` cells " of the phase space lattice as shown in
Fig.~\ref{fig PhaseSpaceLatticen30}(a). The center of each cell is
a stable point corresponding to either a local minimum or a local
maximum of $g$ (see more details in section C of the Appendix).
The area inside the cell represents the basin of attraction for
the stable state in the center. In Fig.~\ref{fig
PhaseSpaceLatticen30}(b), we show the radial structure of the
quasienergy $g$ by plotting it along two angular directions
$\theta=0$ and $\theta=\pi/n$. We see the quasienergy oscillates
as a function of the radius $r$ in the form of Bessel functions
$J_n(r)$. We divide the whole lattice structure into `` loops ",
which correspond to ring-like areas in Fig.~\ref{fig
PhaseSpaceLatticen30}(a) between two radii which satisfy
$J_n(r)=0$. We label them from inside to outside by Roman numerals
$\Rmnum{1}$, $\Rmnum{2}$, $\Rmnum{3}$ and so on as indicated in
Fig.~\ref{fig PhaseSpaceLatticen30}(b).

\section{Quasinumber theory}

\begin{figure}[tp]
\begin{tabular}{lll}
    \textbf{(a)} & \textbf{(b)}& \\
    \includegraphics[scale=0.65]{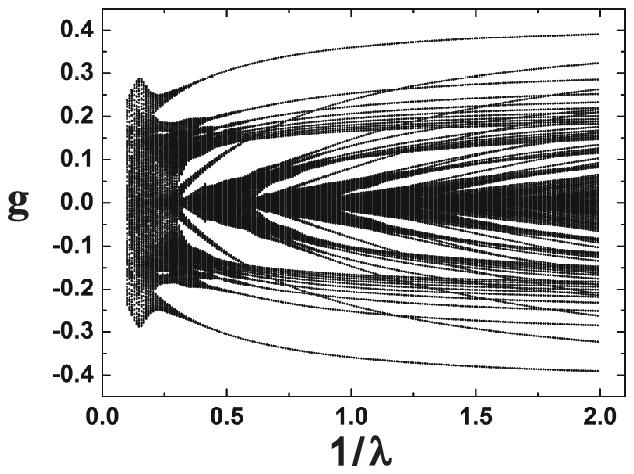} &
    \ \ \ \ \ \ \ \ \ \ \ \ \ \includegraphics[scale=0.65]{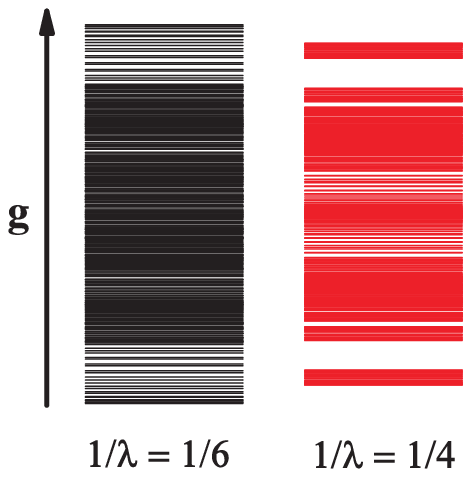}\\
     \textbf{(c)} & \textbf{(d)}& \\
    \includegraphics[scale=0.65]{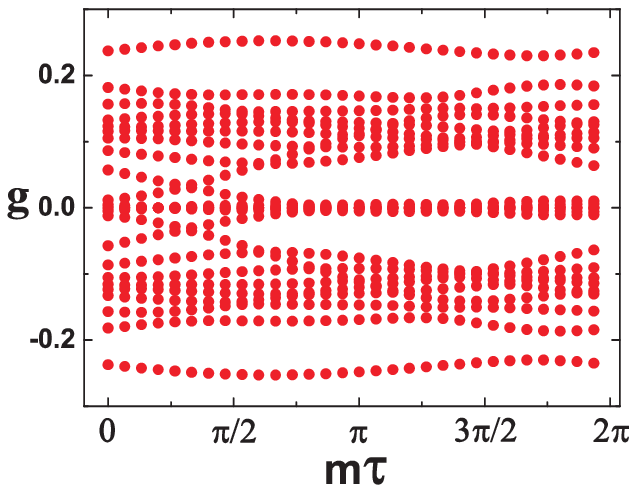} &
    \includegraphics[scale=0.65]{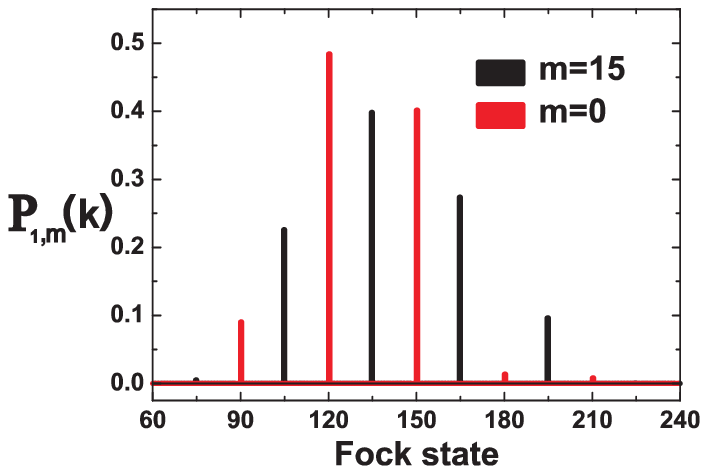}
\end{tabular}
\caption{\footnotesize{\textbf{Quasienergy band structure}: a)
Quasienergy spectrum of Hamiltonian (\ref{HRWA}) as function of
parameter $1/\lambda$. The gap closes at $\lambda\approx 5$. b)
Gapless quasienergy spectrum with parameter $\lambda=6$ and gapped
spectrum $\lambda=4$. Parameter $m$ is the quasinumber of each
state. c) The quasienergy band structure in the reduced Brillouin
zone with $\lambda=4$. The parameter $\tau=2\pi/n$ is the angular
lattice constant. d) Probability distribution over Fock states of
quasinumber states in the bottom band with $m=0$ (red) and $m=15$
(black) respectively. }} \label{fig QuasienergySpectrum}
\end{figure}

We diagonalize the quantum Hamiltonian (\ref{HRWA}) and study the
properties of its quasienergy spectrum. With zero detuning
$\delta\omega=0$, and driving $\mu=1$, the spectrum is only
determined by the effective Planck constant $\lambda$. In
Fig.~\ref{fig QuasienergySpectrum}(a) we show the structure of the
quasienergy spectrum as function of the parameter $1/\lambda$. It
is clear that the quasienergy spectrum is symmetric with respect
to $g = 0$ because of the chiral symmetry. We also see that gaps
in the spectrum are opened for small $\lambda$ and disappear for
sufficiently large $\lambda$. The transition happens around
$\lambda\approx5$. We will calculate the gaps using WKB theory and
discuss the physical mechanism of gap closing below.

In Fig.~\ref{fig QuasienergySpectrum}(b) we show the gapless
quasienergy spectrum for $\lambda=6$ and the band structure of the
spectrum for $\lambda=4$. The band structure comes from the
discrete phase translation symmetry. We introduce the quasinumber
theory \cite{Guo} according to Bloch's theorem. Due to
$\hat{T}_\tau^\dagger \hat{g} \hat{T}_\tau = \hat{g}$ for
$\tau={2\pi}/{n}$, the eigenstates $\psi_m(\theta)$ of the
quasienergy Hamiltonian,
$\hat{g}\psi_m(\theta)=g(m)\psi_m(\theta)$,
 have the form $
\psi_m(\theta)=\varphi_m(\theta)e^{-im\theta}$, with a periodic
function $\varphi_m(\theta+\tau)=\varphi_m(\theta)$. Here, the
integer number $m$ is called \textit{quasi-number}, which is
conjugate to the phase $\theta$. It is an analogue of the
quasi-momentum $\overset{\rightharpoonup}{\vphantom{a}\smash k}$
in a crystal. In Fig.~\ref{fig QuasienergySpectrum}(c), we plot
the quasienergy band structure in the reduced Brillouin zone
$m\tau\in[0,2\pi)$. We count the bands  from the bottom and
relabel the eigenstates $\psi_m(\theta)$ by $\psi_{l,m}(\theta)$,
where the subscript $l=1,2,...$ indicates the band that the
eigenstate belongs to.

In Fig.~\ref{fig QuasienergySpectrum}(d), we plot the occupation
number statistics of Fock states, i.e.,
$\mathrm{P_{1,m}}(k)=|\langle k|\psi_{1,m}(\theta)\rangle|^2$, for
quasinumber states with $m=0$ and $m=15$ in the first band $l=1$.
As we can see from the probability distribution, the
quasinumber states are the superposition of Fock states with
photon numbers being multiples of $n$. To visualize the
quasinumber states, we plot the $Q$-function of state
$\psi_{1,0}(\theta)$ in Fig.~\ref{fig PhaseSpaceLatticen30}(c).
The $Q$-function is a quasi-probability distribution in phase
space \cite{Scully} defined by $Q(\alpha,\alpha^*)\equiv|\langle
\alpha|\psi_{l,m}(\theta)\rangle|^2/ \pi$, where $|\alpha\rangle$
is the coherent state given by
$\hat{a}|\alpha\rangle=\alpha|\alpha\rangle$ or
$|\alpha\rangle=e^{-|\alpha|^2/2}\sum_{k=0}^\infty
\alpha^k/\sqrt{k!}|k\rangle$. The crystalline structure of
$Q$-function in angular direction reflects the $n$-fold discrete
phase translation symmetry.

From the form of the Q-function we see that the
eigenstates of the system $\psi_m(\theta)$ are delocalized states in
phase space, which are superposition of localized states
corresponding to the discrete energy levels as indicated in
Fig.~\ref{fig PhaseSpaceLatticen30}(b). We label these levels in
the first loop by $Level\ \Rmnum{1}-1$, $Level\ \Rmnum{1}-2$ and
those in the second loop $Level\ \Rmnum{2}-1$ etc. In the
semiclassical limit, these quantum levels become classical orbits
of iso-quasienergy contours represented by the boundaries of the
colored elliptical areas inside each cell as shown in
Fig.~\ref{fig PhaseSpaceLatticen30}(a). The shapes of these orbits
vary in different loops as displayed on the top of Fig.~\ref{fig
TunnellingAndThermal}(d).

\section{Quasienergy band structure}

The formation of quasienergy bands near the bottom can be
understood in the frame of the tight-binding model. If we neglect
quantum tunnelling, the $n$ localized states in each loop are $n$
degenerate states. If we consider quantum tunnelling, they are
broadened and form bands. We can label the bands by the labels of
corresponding localized levels, e.g., the bottom band of the whole
quasienergy spectrum is $Band\ \Rmnum{1}-1$. We can describe the
structure of the $l$-th tight-binding band approximately by
\begin{eqnarray}\label{spectruminoneband}
g_l(m)=E_l-2|J_l|\cos(m\tau+\delta_l\tau).
\end{eqnarray}
Here, $E_l$ represents the center of the $l$-th band and the
quasienergy of the corresponding localized level. The $l$-th bandwidth
$d_l$ is determined by the tunnelling rate, i.e., $d_l=4|J_l|$.
From Fig.~\ref{fig QuasienergySpectrum}(c) we see that the bands
are not symmetric with respect to the center of the Brillouin zone in
general. We describe the asymmetry by an \textit{asymmetry factor}
$\delta_l$. The asymmetry factor comes from the fact that the two
dimensions of phase space are not commutative. We will calculate
the gaps, bandwidths and asymmetry factor by WKB theory
below.

\subsection{ Quantum tunnelling in phase space}

From the commutation relation (\ref{comm}), it can be shown that
$[\hat{r}^2/2,\hat{\theta}]\approx i\lambda$ in the region of
$r\gg
   1$ \cite{Guo}.  We can view operators $\hat{r}^2/2$ and $\hat{\theta}$ as ``coordinate" and ``momentum" respectively, i.e.,
$\hat{\theta}\approx-i\lambda r^{-1}\partial/\partial r$. In the
semiclassical limit, the variables $r^2/2$ and $\theta$ define the
phase space for our WKB calculation. In Fig.~\ref{fig
TunnellingAndThermal}(a), we plot the quasienergy $g$ in the range
of $\theta\in[-2\pi/n,2\pi/n]$. For a fixed $g$, all the branches
of classical orbits are given by
\begin{eqnarray}\label{WKBorbit}
\theta_{\pm}(r,g)=\frac{\pi}{2}\pm\frac{1}{n}\arccos\Big[\frac{g-\epsilon
r^2/2}{2\mu J_n(r)}\Big]+k\frac{2\pi}{n},
\end{eqnarray}
where $k$ takes integers $0$, $1$, $2$, $\cdot\cdot\cdot$, and
$n-1$. Two real solutions $\theta_{\pm}(\xi,g)$ together represent
one closed classical orbit. There are $n$ identical orbital
branches with only a ${2\pi}/{n}$-shift of $\theta$. From the
condition $|(g-\epsilon r^2/2)/[2\mu J_n(r)]| < 1$, we can
determine the boundaries of classical motion. In Fig.~\ref{fig
TunnellingAndThermal}(a), we indicate the boundaries of classical
motion by $r_1^2/2$, $r_2^2/2$ and $r_3^2/2$ in the phase space
spanned by $r^2/2$ and $\theta$. The region between $r_2^2/2$ and
$r_3^2/2$ is the classically forbidden region for the fixed
quasienergy $g$. In the quantum regime, however, the states can
tunnel into each other. In Fig.~\ref{fig TunnellingAndThermal}(a),
we show how the two neighboring $Level\ \Rmnum{1}-1$ states tunnel
into each other through phase space. The main tunnelling path with
least action is indicated by the white arrows in the same plot.
The optimal path is  to tunnel first into the nearest region in
$Loop\ \Rmnum{2}$ across one saddle point (white dot) and then
tunnel back to the neighboring $Level\ \Rmnum{1}$-1 across another
saddle point. There also exist many other possible tunnelling
paths in phase space, e.g., the path indicated by yellow arrows in
Fig.~\ref{fig TunnellingAndThermal}(a). But the contributions from
these paths are exponentially small compared to the main
tunnelling path (see more details in section B of the Appendix).

\subsection{Quasienergy levels and bandwidths}

From the WKB theory, we know the phase space area enclosed by the
classical orbit is quantized according to the so called
Bohr-Sommerfeld quantization condition \cite{Landau}
\begin{eqnarray}\label{Bohr-Sommerfeld}
S(g)&=&\frac{2}{n}\int_{r_1}^{r_2}\arccos\Big[\frac{g-\epsilon
r^2/2}{2\mu J_n(r)}\Big]\ rdr\nonumber\\
&=&2\pi\lambda(k+\frac{1}{2}),
\end{eqnarray}
where $k$ takes nonnegative integers. From the above condition we
can calculate the quasienergy levels. As shown in Fig.\ref{fig
TunnellingAndThermal}(b), the left subfigure shows several lowest
levels calculated using the quantization condition
(\ref{Bohr-Sommerfeld}). We compare our WKB calculation to the
numerical simulation. The agreement is very good. Noticeably,
$Level\  \Rmnum{1}$-2 and $Level\ \Rmnum{2}$-1 cross each other
near $\lambda=1.2$. The level crossing has significant effect on
the bandwidths as we discuss below.

The width of the $l$-th band $d_l$ is given by the tunnelling rate $J_l$,
i.e., $d_l=4|J_l|$. The amplitude of $J_l$ is given by the
integral of the imaginary part of ``momentum" $\theta$ in the
classical forbidden region $r_2<r<r_3$
\begin{eqnarray}\label{AbsJ}
|J_l|=\frac{\lambda}{2\pi}\Big(\frac{dS}{dg}\Big)^{-1}\Big|_{g=g_l}\exp{\Big(-\frac{2}{\lambda}\int_{r_2}^{r_3}\mathrm{Im}[\theta]\
 rdr\Big)}.
\end{eqnarray}
Here, $S(g)$ in the prefactor as function of $g$ is given by the
first equality of Eq.(\ref{Bohr-Sommerfeld}). In section B of the
Appendix, we give a detailed description of the behavior of
$\mathrm{Im}[\theta]$ in the classical forbidden region. Here we
just present our results. In Fig.~\ref{fig
TunnellingAndThermal}(b) we show the bandwidths of $Level\
\Rmnum{1}$-1 and $Level\ \Rmnum{1}$-2 calculated by
Eq.(\ref{AbsJ}) and compare them to the numerical calculation.
There is a cusp in the curve of $Level\ \Rmnum{1}$-2. This happens
because of the crossing of $Level\ \Rmnum{1}$-2 and $Level\
\Rmnum{2}$-1 which significantly enhances the quantum tunnelling
of $Level\ \Rmnum{1}$-2. In this case, we need to consider three
interacting levels, i.e., two neighboring  $Level\ \Rmnum{1}$-2
states and the medium state of $Level\ \Rmnum{2}$-1 as indicated
by the closed orbits in Fig.~\ref{fig TunnellingAndThermal}(a).
The Hamiltonian of three interacting levels (TIL) is described by
the following $3 \times 3$ matrix
\[ H_{TIL}=\left( \begin{array}{ccc}
g_{1} & J_{12} & J_{11} \\
J_{12} & g_{2} & J_{12} \\
J_{11} & J_{12} & g_{1} \end{array} \right). \] Here $g_{1}$,
$g_{2}$ represent the quasienergies of $Level\  \Rmnum{1}$-2 and
$Level\  \Rmnum{2}$-1 respectively. Parameter $J_{11}$ represents
the tunnelling rate between the two neighboring  $Level\
\Rmnum{1}$-2 states. Parameter $J_{12}$ represents the tunnelling
rate between the state of $Level\ \Rmnum{1}$-2 and the state of
$Level\ \Rmnum{2}$-1. The tunnelling rate $J_{11}$ is given by
Eq.(\ref{AbsJ}) by taking $g=g_1$, while the tunnelling rate
$J_{12}$ is given by
\begin{eqnarray}\label{AbsJ21}
J_{12}=\frac{\lambda}{2\pi}\Big(\frac{dS}{dg}\Big)^{-1}\Big|_{g=g_2}\exp\Big(-\frac{1}{\lambda}\int_{r_2}^{r_3}\mathrm{Im}[\theta]\
rdr\Big).
\end{eqnarray}
We can get the modified quasienergy levels by diagonalizing the
matrix $H_{TIL}$. The level spacing $\Delta_{11}$ of the two
modified $Level\ \Rmnum{1}$-2 states gives the effective
tunnelling rate between them. Therefore, the correct bandwidth of
$Band\  \Rmnum{1}$-2 is $2\Delta_{11}$.

\subsection{Band asymmetry and artificial magnetic field}

From Fig.~\ref{fig QuasienergySpectrum}(c), we see that the
quasienergy bands are not symmetric with respect to the center of
the reduced Brillouin zone. The asymmetry is described by the
asymmetry factor $\delta_l$. In the frame of tight-binding
approximation, the Bloch eigenstate $\psi_{lm}(\theta)$ is given
by
$\psi_{lm}(\theta)=1/\sqrt{n}\sum_{q=0}^{n-1}e^{imq\tau}{\hat{T}_\tau^q}\phi_l(\theta)$,
where $\phi_l(\theta)$ is the localized wave functions forming the
band. The quantum tunnelling rate can be calculated by
$J_l=-\int[\hat{T}_\tau\phi_l(\theta)]^*\hat{g}\phi_l(\theta)$.
The corresponding quasienergy spectrum of the $l$-th band then is
$g_l(m)=\int_0^{2\pi}\psi^*_{lm}(\theta)\hat{g}\psi_{lm}(\theta)d\theta\approx
E_l-J_le^{im\tau}-J_l^*e^{-im\tau}.
$
The band asymmetry comes from the fact that quantum tunnelling
rate $J_l$ in driven systems is generally a complex number
\cite{Guo,MarthalerTunnelling}, i.e.,
$J_l=|J_l|e^{-i\delta_l\tau}$, and the phase parameter $\delta_l$
is exactly the asymmetry factor. We can calculate the phase
$\delta_l$ using the WKB theory we developed above.

In fact, when $r$ is approaching one of the roots $r^{(0)}$ with
$J_n(r^{(0)})=0$, from Eq.(\ref{WKBorbit}) we see the amplitude of
``momentum" $\theta$ goes to infinity
$|\theta(r^{(0)})|\rightarrow \infty$. This means the WKB
approximation breaks down near the  root of the Bessel function
$J_n(r^{(0)})=0$ and we need a connecting condition. Because
$r^{(0)}\gg 1$, we can expand the phase translation operator
$\hat{T}_\tau=e^{-i\tau \hat{a}^\dag \hat{a}}$ by \cite{Guo}
$\hat{a}^\dag \hat{a}\approx
\lambda^{-1}(r^{(0)})^2/2+i\partial/\partial \theta$ and the
connecting condition, i.e., the neighboring localized state of
$\phi_l(\theta)$, is given by $\hat{T}_\tau\phi_l(\theta)\approx
e^{-i\lambda^{-1}(r^{(0)})^2\tau/2}\phi_l(\theta+\tau)$. Thus we
get the symmetry factor
$\delta_l=\delta_l^0+\lambda^{-1}(r^{(0)})^2/2$, where
$\delta_l^0$ is the residual asymmetry beyond WKB calculation and
can be removed by redefining the phase translation operator
$\hat{T}_\tau=e^{-i\tau {(\hat{a}^\dagger} \hat{a}-\delta_l^0)}$.
The asymmetry factor $\delta_l$ is linearly dependent on the
parameter $1/\lambda$ with the slope $(r^{(0)})^2/2$ differing
between bands. If we count $r^{(0)}=0$ as the first root of
$J_n(r)$, then the asymmetry factors of bands in the $l$-th
($l\geq2$) loop are all given by the $l$-th ($l\geq2$) root of the
Bessel function. But the asymmetry factors of the bands in the
first loop are determined by the second root of the Bessel
function. The reason is that the localized states inside the first
loop tunnel through its upper boundary while states in other loops
tunnel through lower boundaries. In section B of the Appendix, we
give more detailed discussion on tunnelling paths and show more
results about the linear relationship between $\delta_l$ versus
$1/\lambda$ for different bands.

The fact that the tunnelling amplitudes are complex means there is
an artificial magnetic field $B_{eff}$ in phase space. Imagine we
have a loop of atoms forming a one dimensional lattice in real
space with magnetic field $B$ across the loop. The magnetic field
induces an additional phase to the tunnelling amplitude between
neighbored atoms $J=|J|e^{-i\delta}$, where $\delta \propto B$ is
called Peierls phase \cite{PerlsPhase}. Comparing the Peierls
phase to the asymmetry factor of the phase space lattice
calculated above, we can identify there is an effective magnetic
field $B_{eff} \propto 1/\lambda$ in phase space. The coordinate
system of a phase space lattice has a noncommutative geometry
\cite{NoncommutativeGeometry}, which is fundamentally different
from spatial lattices. It is this noncommutative phase space which
creates an artificial magnetic field and is responsible for the
asymmetry of the quasienergy band structure.

\begin{figure}[tp]
\begin{tabular}{lll}
    \textbf{(a)} & \textbf{(b)}& \\
    \includegraphics[scale=0.42]{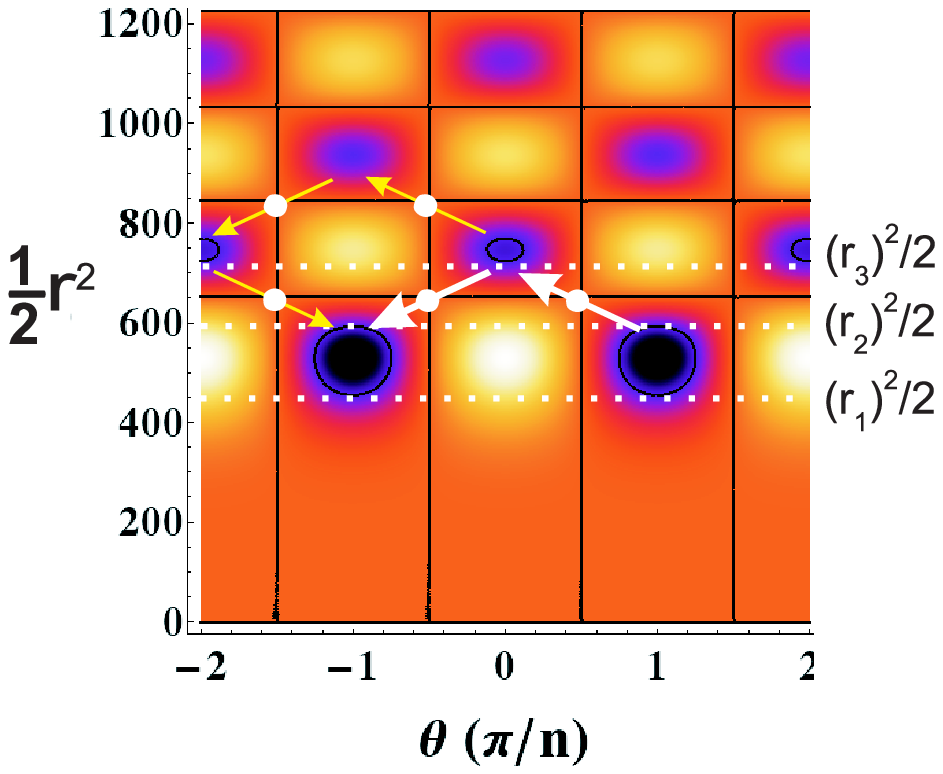} &
    \includegraphics[scale=0.42]{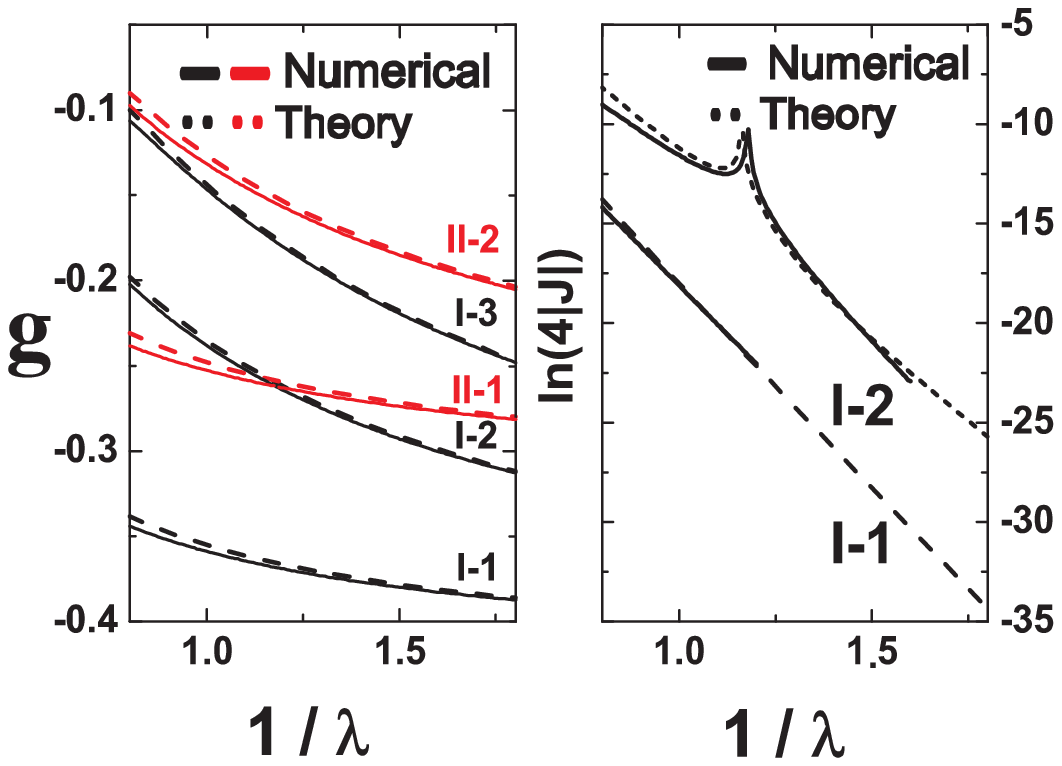}\\
     \textbf{(c)} & \textbf{(d)}& \\
    \includegraphics[scale=0.42]{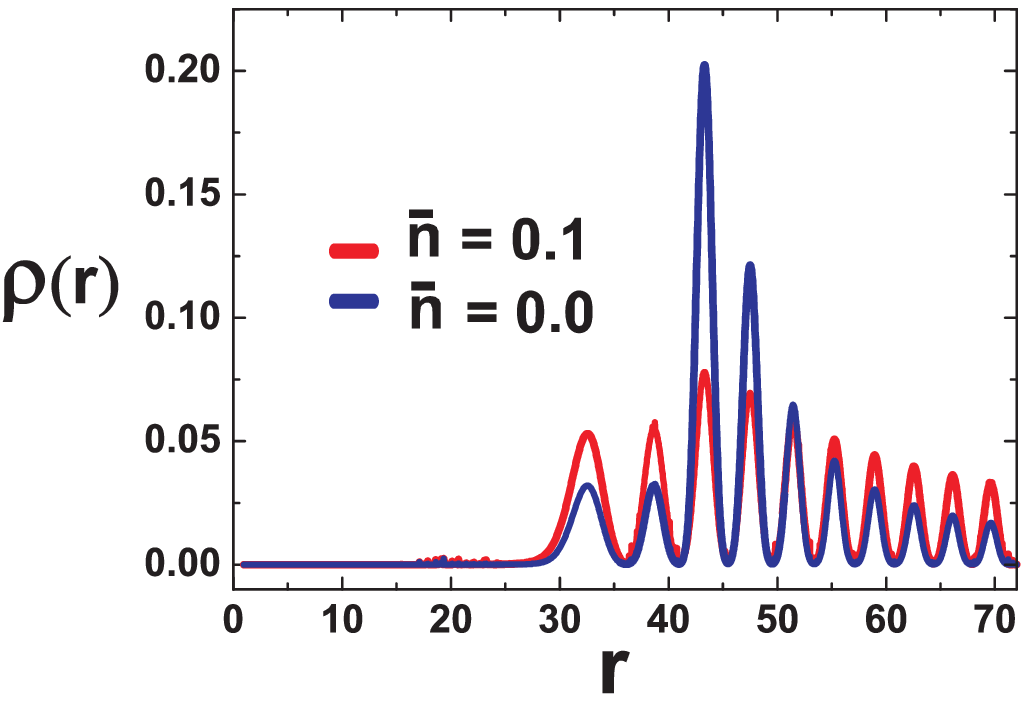} &
    \includegraphics[scale=0.42]{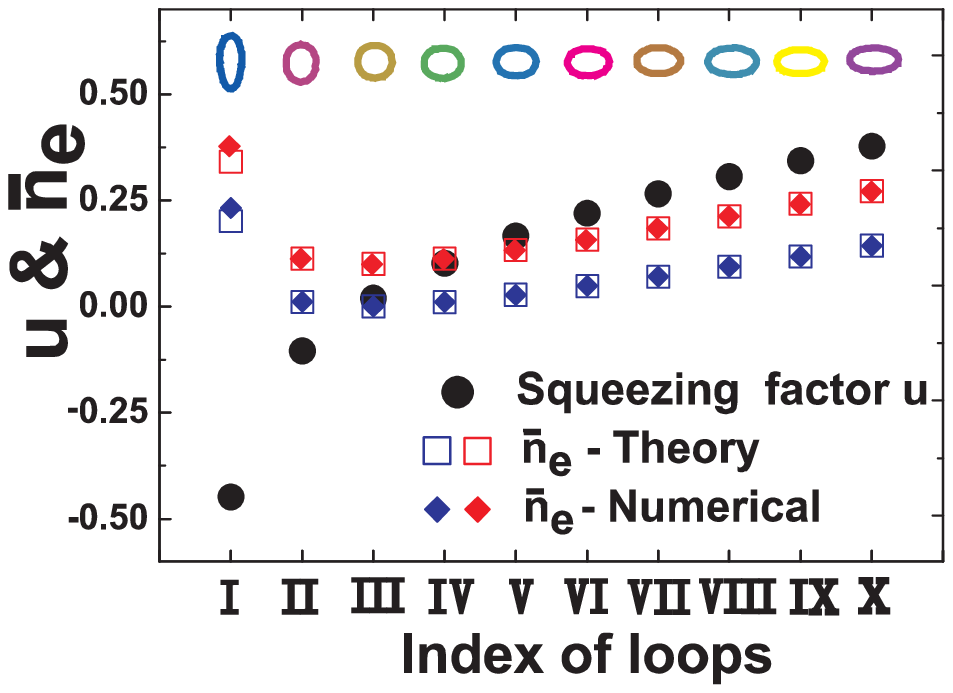}
\end{tabular}
\caption{\footnotesize{\textbf{Quantum tunnelling and Thermal
distribution}: a) Quantum tunnelling processes in phase space
spanned by variables $r^2/2$ and $\theta$. We plot the quasinerngy
$g$ in the interval $\theta\in [-\pi/n,\pi/n]$. White dots are the
unstable saddle points. White arrows indicate the main tunnelling
path of two neighbored $Level\ \Rmnum{1}$-1 states and the yellow
arrows indicate other possible tunnelling paths. b) The
quasienergy levels (left) and bandwidths (right) from WKB
calculation (dashed lines) and numerical simulations (solid
lines). c) Density function of stationary distribution along the
radius $r$ at different temperatures. d) The squeezing factor $u$
and effective temperature $\bar{n}_e$ of the first ten loops of
phase space lattice. The colored circles on the top represent the
shapes of classical orbits near the bottom of each loop, which
reflect the squeezing factor $u$. }} \label{fig
TunnellingAndThermal}
\end{figure}

\section{Dissipative dynamics}

The above calculation of the quasienergy bandstructure does not
consider the dissipative environment. In actual experiments, due
to the quantum and thermal fluctuations, the dynamics in a phase
space lattice is non-unitary. For a driven system, we can measure
the non-equilibrium stationary state in experiments. We use the
master equation method to describe the dissipative evolution
caused by thermal and quantum fluctuations. Already previously it
has been shown that a Lindblad type of master equation
\cite{ME1Dykman,ME2Dykman,MEGuo,MEZoller} is sufficient as
description,
\begin{equation}\label{ME}
\frac{\partial \rho}{\partial
t}=-\frac{i}{\lambda}[\hat{g},{\rho}]
 + \kappa(1+\bar{n}){\cal D}[a]{\rho} +  \kappa \bar{n}{\cal
 D}[a^{\dagger}]{\rho},
\end{equation}
where the time $t$ is dimensionless and scaled by the natural
frequency $\omega$. The Lindblad superoperator is defined through
${\cal D}[A]{\rho}\equiv A{\rho}A^{\dagger}
 -(A^{\dagger}A{\rho}+{\rho}A^{\dagger}A )/2 $, $\bar{n}=(e^{\hbar\omega_0/k_BT}-1)^{-1}$
is the Bose distribution and $\kappa$ is the dimensionless damping
also scaled $\omega$.

Based on the master equation (\ref{ME}), we calculate the density
matrix of the stationary distribution in the basis of the Fock
states $\{|k\rangle, k=0,1,\cdot\cdot\cdot\}$. By the relationship
of $k=r^2/(2\lambda)$, we can find the propbability density along
a circle with radius $r$, i.e., $\rho(r)=r\lambda^{-1}\langle k
|\rho| k\rangle$. In Fig~\ref{fig TunnellingAndThermal}(c), we
plot $\rho(r)$ for different temperatures $\bar{n}=0$ and
$\bar{n}=0.1$. We see that $\rho(r)$ oscillates with radius $r$.
The zero nodes of $\rho(r)$ actually correspond to the boundaries
of phase space lattice loops. Because the quantum heating
\cite{HeatingDykman} of each loop is not the same, the
probabilities over the loops are not equally distributed. On the
bottom of each loop, the stationary distribution can be described
by an effective temperature $\bar{n}_e$. The localized ground
state of each loop can be approximately described by a squeezed
state with the squeezing factor $u$ and the corresponding
effective temperature is given by
$\bar{n}_e=|u|^2+\bar{n}(2|u|^2+1)$ (see more details in section D
of the Appendix). In our case, as we can see from Fig~\ref{fig
TunnellingAndThermal}(c), the peak of $\rho(r)$ is in the third
loop. The reason is the effective temperature of the third loop is
lower than other loops. In Fig~\ref{fig TunnellingAndThermal}(d),
we calculate the squeezing factor $u$ and the effective
temperature $\bar{n}_e$ for the first ten loops and compare them
to fully numerical simulations. The agreement is very good.
 Another interesting fact is the squeezing factor
$u$ changes from a negative value to a positive value. This means
the shape of the squeezed state in each loop is different as
displayed by the colored orbits on the top of Fig~\ref{fig
TunnellingAndThermal}(d). The orbital shapes are taken from the
plot in Fig.~\ref{fig PhaseSpaceLatticen30}(a). The third orbit is
very close to a round circle, which means the squeezing factor
$u\approx 0$ and the resulting effective temperature
$\bar{n}_e \approx \bar{n}$. The stationary distribution can be
directly measured in the experiments \cite{ColdAtoms}.

\section{Discussion}

The phase space lattice can also realized in circuit-QED systems,
i.e., a superconducting cavity coupled to Josephson junctions. The
Hamiltonian is $H_{cQED}=\hbar\omega a^\dagger a + 2E_J\cos(4\pi
e\hbar^{-1}\Phi)\cos\varphi$. The Josephson junction can be driven
by either a dc voltage \cite{AnkerholdJJOsci,ArmourJJOsci}, which
creates $\varphi=\varphi_0+\omega_d t$ with $\omega_d =2eV/\hbar$,
or a time-dependent magnetic flux \cite{cQEDJJSchoen}
$\Phi=\omega_d t/(4\pi e\hbar^{-1})$ . The effective Planck
constant in this case is $\lambda =8\pi\omega L / (h/e^2)$, where
$L$ is the inductance of the circuit and $h/e^2\approx 25.8$
$k\Omega$ is the von Klitzing constant.  The typical impedance
$\omega L$ of circuit-QED systems using only geometrical inductors
and capacitors, can not exceed the characteristic impedance of
vacuum $\mu_0 c\approx376.73$ $\Omega$ \cite{SuperInductance1},
which means that we have $\lambda < 0.015$ in circuit-QED systems.
However, there are several proposals to realized a
super-inductance based on the design of Josephson junction arrays
\cite{SuperInductance1,SuperInductance2} which can increase the
impedance significantly up to $35$ $k\Omega$ resulting
$\lambda>1$. Thus, it is possible to realize phase space lattices
in circuit-QED systems combined with a proper design of Josephson
junction arrays.

\bigskip

\textbf{Acknowledgements}

 We acknowledge helpful discussions with Dr. P. Kotetes, Dr. P. Jin, Prof. G. Johansson, Prof. G. Sch\"on,
Prof. F. Marquart and Dr. V. Peano.

\section*{Supplemental Material}

\subsection{Hamiltonians}

In this section, we give detailed derivation from the
time-dependent Hamiltonian (1) to the RWA Hamiltonian (2) and the
semiclassical Hamiltonian (4) in the main text. To be convenient,
we write the original Hamiltonian of ultracold in the driven
optical lattice atoms here  again
\begin{equation}\label{DrivenOpticalLattice}
H_{DOL}=\frac{p^2}{2m}+\frac{1}{2}m\omega^2x^2+2A\cos(kx+\omega_dt).
\end{equation}
Now, we introduce $a,\ a^\dagger$ via
$x=\sqrt{\hbar/(2m\omega)}(a^\dagger+a)$ and
$p=i\sqrt{m\hbar\omega/2}(a^\dagger-a)$.  By introducing parameter
$\lambda\equiv\hbar k^2/(m\omega)$, we map the Hamiltonian
(\ref{DrivenOpticalLattice}) to the following
\begin{equation}\label{HJJcavity}
H(t)=\hbar\omega a^\dagger a +
2A\cos\Big[\sqrt{\frac{\lambda}{2}}(a^\dagger+a)+\omega_dt\Big].
\end{equation}
We introduce the scaled coordinate and momentum operators
$\hat{Q}=\sqrt{\frac{\lambda}{2}}(a^\dagger+a)$ and
$\hat{P}=i\sqrt{\frac{\lambda}{2}}(a^\dagger-a)$ with the
noncommutative relationship $[ \hat{Q},\hat{P} ]=i\lambda.$ We
write Hamiltonian (\ref{HJJcavity}) in an alternative form
\begin{equation}\label{HQP}
H(t)=\frac{1}{2\lambda}\hbar\omega (\hat{Q}^2+\hat{P}^2) +
2A\cos(\hat{Q}+\omega_dt),
\end{equation}
Now, we employ an unitary operator
$U=e^{i\hbar\frac{\omega_d}{n}a^\dagger at}$ to transform
Hamiltonian (\ref{HQP}) into a rotating frame with frequency
$\omega_d/n$
\begin{eqnarray}\label{Hrf}
H_{RF}&=&UH(t)U^\dagger-iU\dot{U}^\dagger\nonumber\\
&=&\frac{1}{2\lambda}\hbar\delta\omega (\hat{Q}^2+\hat{P}^2)+2AU\cos(\hat{Q}+\omega_dt)U^\dagger\nonumber\\
&=&\frac{1}{2\lambda}\hbar\delta\omega
(\hat{Q}^2+\hat{P}^2)+(Ae^{i\omega_d
t}Ue^{i\hat{Q}}U^\dagger+h.c.)\nonumber\\
&=&\frac{1}{2\lambda}\hbar\delta\omega
(\hat{Q}^2+\hat{P}^2)+\{Ae^{i\omega_d
t}e^{i[\hat{Q}\cos(\omega_dt/n)+\hat{P}\sin(\omega_dt/n)]}+h.c.\}\nonumber\\
&\equiv&\frac{1}{2\lambda}\hbar\delta\omega
(\hat{Q}^2+\hat{P}^2)+[Ae^{i\omega_d t}M(\hat{Q},\hat{P})+h.c.].
\end{eqnarray}
Here, we define $M(\hat{Q},\hat{P})\equiv
e^{i[\hat{Q}\cos(\omega_dt/n)+\hat{P}\sin(\omega_dt/n)]}$ and the
detuning $\delta\omega\equiv\omega_0-\omega_d/n$. To calculate the
matrix element of $M(\hat{Q},\hat{P})$, we define the displacement
operator $D(\alpha,\alpha^*)$ by
\begin{eqnarray}\label{D}
D(\alpha,\alpha^*)&\equiv& \exp\Big(\alpha a^\dagger - \alpha^*
a\Big)\nonumber\\
&=& \exp\Big(\mathrm{Re}[\alpha] (a^\dagger-a) + i\
\mathrm{Im}[\alpha](a^\dagger+a)\Big).
\end{eqnarray}
Since the operator $M(\hat{Q},\hat{P})$ can be written as
\begin{eqnarray}\label{}
M(\hat{Q},\hat{P})&\equiv&\exp\Big[{i\ \Big(\hat{Q}\cos(\omega_dt/n)+\hat{P}\sin(\omega_dt/n)\Big)}\Big]\nonumber\\
&=&\exp\Big[-\sqrt{\frac{\lambda}{2}}\sin(\omega_dt/n)(a^\dagger-a)\nonumber\\
&&+i\sqrt{\frac{\lambda}{2}}\cos(\omega_dt/n)(a^\dagger+a)\Big],
\end{eqnarray}
we get the relationship between the parameter $\alpha$ of
$D(\alpha,\alpha^*)$ and parameters of $M(\hat{Q},\hat{P})$
\begin{eqnarray}\label{relationship}
\alpha&\equiv&-\sqrt{\frac{\lambda}{2}}\sin(\omega_dt/n)+i\sqrt{\frac{\lambda}{2}}\cos(\omega_dt/n)\nonumber\\
&=&\sqrt{\frac{\lambda}{2}}e^{i(\varphi+\pi/2)},
\end{eqnarray}
with $\varphi=\omega_dt/n$. We further define the following
notations
\begin{eqnarray}\label{}
\mathrm{Coherent\ \  state}&:&\ \ |\alpha\rangle\equiv
e^{-\frac{1}{2}|\alpha|^2}\sum_{k=0}^{\infty}\frac{\alpha^k}{\sqrt{k!}}|k\rangle,\nonumber\\
&&\ \ \langle\beta|\alpha\rangle=e^{\alpha\beta^*-(|\alpha|^2+|\beta|^2)/2},\nonumber\\
\ \nonumber\\
\mathrm{Displaced \ \ Fock\ \  state}&:&\ \
|\alpha,k\rangle\equiv
D(\alpha,\alpha^*)|k\rangle,\nonumber\\
&&\ \ |0,k\rangle=|k\rangle.
\end{eqnarray}
According to Eq.(3.11) in the Ref. of \textit{Quantum Opt. 3, 359
(1991)}\cite{Wunsche}, we have
\begin{eqnarray}\label{}
\langle\beta,l|\alpha,k\rangle=\langle\beta|\alpha\rangle\sqrt{\frac{l!}{k!}}(\beta^*-\alpha^*)^{k-l}L_l^{k-l}\Big\{\Big|\beta-\alpha\Big|^2\Big\}.
\end{eqnarray}
Here, $L_l^{k-l}(\bullet)$ is the Laguerre polynomials. Let
$\beta=0$, we have the exact form of matrix element of
displacement operator $D(\alpha,\alpha^*)$
\begin{eqnarray}\label{}
\langle l|\alpha,k\rangle&\equiv&\langle
l|D(\alpha,\alpha^*)|k\rangle\nonumber\\
 &=&e^{-|\alpha|^2/2+i\pi
(k-l)}(\alpha^*)^{k-l}\sqrt{\frac{l!}{k!}}L_l^{k-l}(|\alpha|^2).
\end{eqnarray}
Using the relationship (\ref{relationship}) we get the explicit
form of matrix elements of $M(\hat{Q},\hat{P})$
\begin{eqnarray}\label{MLaguerre}
\langle l|M(\hat{Q},\hat{P})|k\rangle
=e^{-\lambda/4+i(k-l)(\pi/2-\omega_dt/n)} \sqrt{\frac{l!}{k!}}\
\Big(\frac{\lambda}{2}\Big)^{\frac{k-l}{2}}L_l^{k-l}(\lambda/2).\nonumber\\
\end{eqnarray}
Thus, quantum Hamiltonian (\ref{Hrf}) is
\begin{eqnarray}\label{}
H_{RF}&=&\frac{1}{2\lambda}\hbar\delta\omega
(\hat{Q}^2+\hat{P}^2)+[Ae^{i\omega_d
t}M(\hat{Q},\hat{P})+h.c.]\nonumber\\
&=&\hbar\delta\omega (a^\dagger a+\frac{1}{2})+[Ae^{i\omega_d
t}M(\hat{Q},\hat{P})+h.c.]\nonumber\\
&=&\hbar\delta\omega (a^\dagger
a+\frac{1}{2})+A\Big[\sum_{k,l}\langle
l|M(\hat{Q},\hat{P})|k\rangle e^{i\omega_d t}|l\rangle\langle
k|+h.c.\Big]\nonumber\\
&=&\hbar\delta\omega (a^\dagger
a+\frac{1}{2})+A\Big[\sum_{k,l}e^{i\omega_d
t}e^{-\lambda/4+i(k-l)(\pi/2-\omega_dt/n)}\nonumber\\
&&\times \sqrt{\frac{l!}{k!}}\
\Big(\frac{\lambda}{2}\Big)^{\frac{k-l}{2}}L_l^{k-l}(\lambda/2)|l\rangle\langle
k|+h.c.\Big].
\end{eqnarray}
Under RWA, we drop the fast oscillating terms ($k-l\neq n$) and
get RWA Hamiltonian ($k-l= n$)
\begin{eqnarray}\label{EaxctRWA}
&&H_{RWA}\nonumber\\
&=&\hbar\delta\omega (a^\dagger
a+\frac{1}{2})+A\Big[\sum_{l}e^{-\lambda/4+in\pi/2} \nonumber\\
&&\times\sqrt{\frac{l!}{(l+n)!}}\
\Big(\frac{\lambda}{2}\Big)^{\frac{n}{2}}L_l^{n}(\lambda/2)\
|l\rangle\langle l+n|+h.c.\Big]\nonumber\\
&=&\hbar\delta\omega (a^\dagger
a+\frac{1}{2})+A\Big[e^{-\lambda/4+in\pi/2}\Big(\frac{\lambda}{2}\Big)^{\frac{n}{2}}\nonumber\\
&&\times\sum_{l}|l\rangle\langle l+n| \sqrt{\frac{l!}{(l+n)!}}\
L_l^{n}(\lambda/2)+h.c.\Big]\nonumber\\
&=&\hbar\delta\omega (a^\dagger
a+\frac{1}{2})+A\Big[e^{-\lambda/4-in\pi/2}\Big(\frac{\lambda}{2}\Big)^{-\frac{n}{2}}\nonumber\\
&&\times\sum_{l}|l\rangle\langle
l+n| \sqrt{\frac{(l+n)!}{l!}}\ L_{l+n}^{-n}(\lambda/2)+h.c.\Big]\nonumber\\
&=&\hbar\delta\omega (a^\dagger
a+\frac{1}{2})+A\Big[e^{-\lambda/4-in\pi/2}\Big(\frac{\lambda}{2}\Big)^{-\frac{n}{2}}\nonumber\\
&&\times\sum_{l}|l\rangle\langle
l|a^n\ L_{a^\dagger a}^{-n}(\lambda/2)+h.c.\Big]\nonumber\\
&=&\hbar\delta\omega (a^\dagger
a+\frac{1}{2})+A\Big[e^{-\lambda/4-in\pi/2}\Big(\frac{\lambda}{2}\Big)^{-\frac{n}{2}}a^n\
L_{a^\dagger a}^{-n}(\lambda/2)+h.c.\Big].\nonumber\\
\end{eqnarray}
Here we have used the relationship \cite{Wunsche} $
L_l^{n}(x)/L_{l+n}^{-n}(x)=(-x)^{-n}(l+n)!/l!$ for $x>0$. We now
scale the RWA Hamiltonian by $\hbar\omega/\lambda$ and get the
dimensionless Hamiltonian $\hat{g}$
\begin{eqnarray}\label{EaxctRWAh}
\hat{g}&\equiv&\frac{\lambda}{\hbar\omega}H_{RWA}\nonumber\\
&=&\lambda\epsilon(a^\dagger
a+\frac{1}{2})+\mu\Big[e^{-\lambda/4-in\pi/2}\Big(\frac{\lambda}{2}\Big)^{-\frac{n}{2}}a^n\
L_{a^\dagger a}^{-n}(\lambda/2)+h.c.\Big],\nonumber\\
\end{eqnarray}
where the parameters $\epsilon=\delta\omega/\omega$ and
$\mu=\lambda A/(\hbar\omega)$ are the dimensionless detuning and
driving strength respectively.

Using the following asymptotic form of Laguerre polynomials
\cite{Bateman,Szegoe}
\begin{eqnarray}\label{aymptotic}
\lim_{k\rightarrow \infty}L_k^{\alpha}(x/k)=k^\alpha
e^{\frac{x}{2k}}x^{-\alpha/2}J_\alpha(2\sqrt{x}),
\end{eqnarray}
we have the following relationship in the limit of $k,l \gg |k-l|$
for a fixed $k-l$
\begin{eqnarray}\label{aymptotic1}
L_k^{k-l}(\lambda/2)&\approx&
e^{\lambda/4}k^{k-l}(k\lambda/2)^{-(k-l)/2}J_{k-l}(2\sqrt{k\lambda/2})\nonumber\\
&=&e^{\lambda/4}\Big(\frac{\lambda}{2k}\Big)^{-\frac{1}{2}(k-l)}J_{k-l}(\sqrt{2k\lambda}).
\end{eqnarray}
Thus, in the semiclassical limit, i.e., $k,l \rightarrow \infty$
and fixed $k-l$, Eq.(\ref{MLaguerre}) goes to the following
\begin{eqnarray}\label{MLaguerrelimit}
\langle l|M(\hat{Q},\hat{P})|k\rangle
&=&e^{-\lambda/4+i(k-l)(\pi/2-\varphi)} \sqrt{\frac{l!}{k!}}\
\Big(\frac{\lambda}{2}\Big)^{\frac{k-l}{2}}L_l^{k-l}(\lambda/2)\nonumber\\
&\approx& e^{i(k-l)(\pi/2-\varphi)} k^{(k-l)/2}
\sqrt{\frac{l!}{k!}} J_{k-l}(\sqrt{2k\lambda})\nonumber\\
&\approx& e^{i(k-l)(\pi/2-\varphi)} J_{k-l}(\sqrt{2k\lambda}).
\end{eqnarray}
Here, we have used the limit relationship $\sqrt{\frac{l!}{k!}}\
k^{(k-l)/2}\rightarrow 1$. Therefore, we have the RWA Hamiltonian
(\ref{EaxctRWAh}) in the Fock representation
$\hat{g}=\sum_{k=0,l=0}^{\infty}f(k,l)|k\rangle\langle l|$ with
\begin{eqnarray}\label{HFock}
f(k,l)&\approx&\lambda\epsilon(k+\frac{1}{2})\delta_{k,l}
+\mu\Big[e^{in\frac{\pi}{2}}J_{k-l}\Big(\sqrt{\lambda(k+l+1)}\Big)\delta_{l-k,n}+h.c.\Big].\nonumber\\
\end{eqnarray}
We define the radial and angular operators $\hat{r}$ and
$\hat{\theta}$ by $a=e^{-i\theta}\hat{r}/\sqrt{2\lambda}$ and
$a^\dag=\hat{r}e^{i\theta}/\sqrt{2\lambda}$. In the Fock
representation, the operator $e^{i\hat{\theta}}$ is defined by
\begin{eqnarray}\label{}
e^{i\hat{\theta}}=\sum_{k=0}^{\infty}|k\rangle\langle k+1|, \ \ \
\ \ \mathrm{and} \ \ \ \ \
e^{-i\hat{\theta}}=\sum_{k=0}^{\infty}|k+1\rangle\langle k|.
\end{eqnarray}
Using the above relationships, we have the following Hamiltonian
in the semiclassical limit $\lambda\rightarrow 0$
\begin{eqnarray}\label{Classicalg}
\hat{g}\rightarrow g=\frac{1}{2} \epsilon r^2+ 2\mu
J_n(r)\cos(n\theta-\frac{n\pi}{2}).
\end{eqnarray}

\subsection{Quantum tunnelling in phase space}

\begin{figure*}
\centerline{\includegraphics[scale=0.65]{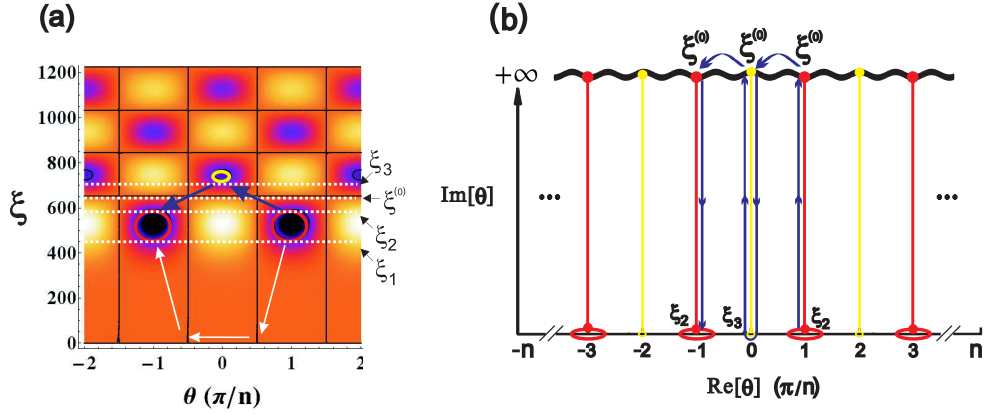}}
\caption{\label{fig TunnellingDiagram}{\bf{Quantum tunnelling in
phase space.}} a) Tunnelling processes of two states in the first
loop. The red closed curves represent two classical orbits in the
first loop. The yellow closed curve represents the classical orbit
in the second loop. Parameters $\xi_1$, $\xi_2$ and $\xi_3$
indicate the classical boundaries of classical motions. Parameter
$\xi^{(0)}$ is the second zero point of Bessel function
$J_n(\sqrt{2\xi})$. The blue arrows indicate the upper tunnelling
path while the white arrows indicate the lower tunnelling path. b)
Tunnelling diagram for calculating tunnelling rate $J_l$. We show
the analytical behavior of $\theta$ in its complex plane. The red
and yellow closed curves on the real axis Re[$\theta$] correspond
to the classical orbits with the same colors in the left figure
(we deviate the orbits slightly from the real axis to illustrate
the shapes of orbits). The red and yellow vertical lines
correspond to the behaviors of imaginary parts Im$[\theta]$ in the
classical forbidden region. The blue curves with arrows indicate
the tunnelling process. The jumping processes happen at the zero
point $\xi^{(0)}$ where the imaginary part Im$[\theta]=+\infty$. }
\end{figure*}

In this section, we give a detailed description about the quantum
tunnelling process in phase space and the analytical behavior of
``momentum" $\theta$ in the complex plane. We also calculate the
asymmetry factor $\delta$ and show its linear relationship with
$1/\lambda$ for different bands. To be convenient, we define a new
variable $\hat{\xi}\equiv\hat{r}^2/2$ here. The semiclassical
Hamiltonian (\ref{Classicalg}) can be rewritten as $g=\epsilon\xi+
2\mu J_n(\sqrt{2\xi})\cos(n\theta-\frac{n\pi}{2})$ in new
variables $\xi$ and $\theta$, which define the ``$\xi-\theta$ "
phase space for our WKB calculation. For a fixed $g$, the general
solutions of classical orbits are
\begin{eqnarray}\label{WKBorbit}
\theta_{\pm}(\xi,g)=\frac{\pi}{2}\pm\frac{1}{n}\arccos\Big[\frac{g-\epsilon\xi}{2\mu
J_n(\sqrt{2\xi})}\Big]+k\frac{2\pi}{n},
\end{eqnarray}
where $k=0,\ 1,\ ,2\ \cdot\cdot\cdot,\ \mathrm{and}\ n-1$
represent the $n$ branches of solutions. Here, we choose the
parameters $\epsilon=0$ and $\mu=-1$. In Fig.~\ref{fig
TunnellingDiagram}, we show three classical orbits for a fixed
$g<0$. The two classical orbits in the first loop are indicated by
red closed curves, which correspond to the following solutions
\begin{eqnarray}\label{WKBorbitRed}
\theta_{\pm}(\xi,g)&=&-\frac{\pi}{n}\pm\frac{1}{n}\arccos\Big[\frac{g-\epsilon\xi}{2\mu
J_n(\sqrt{2\xi})}\Big],\ \ \ \mathrm{and}\nonumber\\
\theta_{\pm}(\xi,g)&=&\frac{\pi}{n}\pm\frac{1}{n}\arccos\Big[\frac{g-\epsilon\xi}{2\mu
J_n(\sqrt{2\xi})}\Big].
\end{eqnarray}
The classical orbit in the second loop is indicated by yellow
closed curve, which corresponds to the following solution
\begin{eqnarray}\label{WKBorbitYellow}
\theta_{\pm}(\xi,g)=\pm\frac{1}{n}\Big(\pi-\arccos\Big[\frac{g-\epsilon\xi}{2\mu
J_n(\sqrt{2\xi})}\Big]\Big).
\end{eqnarray}
In the regime of $\Big|(g-\epsilon\xi)/\Big[2\mu
J_n(\sqrt{2\xi})\Big]\Big| < 1$, two real solutions
$\theta_{\pm}(\xi,g)$ together represent one closed classical
orbit  $\theta(\xi,g)$. In Fig.~\ref{fig TunnellingDiagram}(left),
the boundaries of classical motions are indicated by the white
dashed lines, i.e., $\xi_1$, $\xi_2$ and $\xi_3$. Beyond the
classical boundaries, the value of $\theta(\xi,g)$ has imaginary
part. In Fig.~\ref{fig TunnellingDiagram}(right), we show the
analytical structures of solutions $\theta_{\pm}(\xi,g)$ in the
complex plane. The closed curves on the real axis of $\theta$
represent classical orbits (we deviate the orbits slightly from
the real axis to illustrate the shapes of orbits). There are $n$
identical orbital branches with only a ${2\pi}/{n}$-shift of
Re[$\theta$] for each type of solution.

In the quantum regime, the classical orbits can tunnel into each
other through the classical forbidden region. In Fig.~\ref{fig
TunnellingDiagram}(left), we show the quantum tunnelling process
of the two states in the first loop in phase space. The
corresponding behavior of Im[$\theta$] is depicted in
Fig.~\ref{fig TunnellingDiagram}(right). Starting from the
classical boundary $\xi_2$ to the zero point of Bessel function
$\xi^{(0)}$, the imaginary part Im[$\theta$] increases from zero
to infinite, where it jumps to another branch of solution. Then it
goes back from infinite to zero as $\xi$ changes from $\xi^{(0)}$
to another classical boundary $\xi_3$. After that, Im[$\theta$]
increases again from zero to infinite as $\xi$ goes from $\xi_2$
to $\xi^{(0)}$, where it jumps again to another branch of
solution. Finally, Im[$\theta$] decreases from infinite to zero as
$\xi$ changes from $\xi^{(0)}$ to the classical boundary $\xi_3$.
As we have discussed in the main text, the amplitude of quantum
tunnelling rate $J_l$ is given by the integral of the imaginary
part of ``momentum" $\theta$ in the classical forbidden region
$\xi_2<\xi<\xi_3$
\begin{eqnarray}\label{AbsJ}
|J_l|=\frac{\lambda}{2\pi}\Big(\frac{dS}{dg}\Big)^{-1}\Big|_{g=g_l}\exp{\Big(-\frac{2}{\lambda}\int_{\xi_2}^{\xi_3}\mathrm{Im}[\theta]d\xi\Big)}.
\end{eqnarray}
The tunnelling process can also happen through lower boundary
$\xi_1$ as indicated by the white arrows in Fig.~\ref{fig
TunnellingDiagram}(left). However, the lower path is much longer
than the upper path. Thus, the contribution to $|J_l|$ from the
lower path is exponentially smaller than the contribution from
upper path.

\begin{figure*}
\centerline{\includegraphics[scale=0.8]{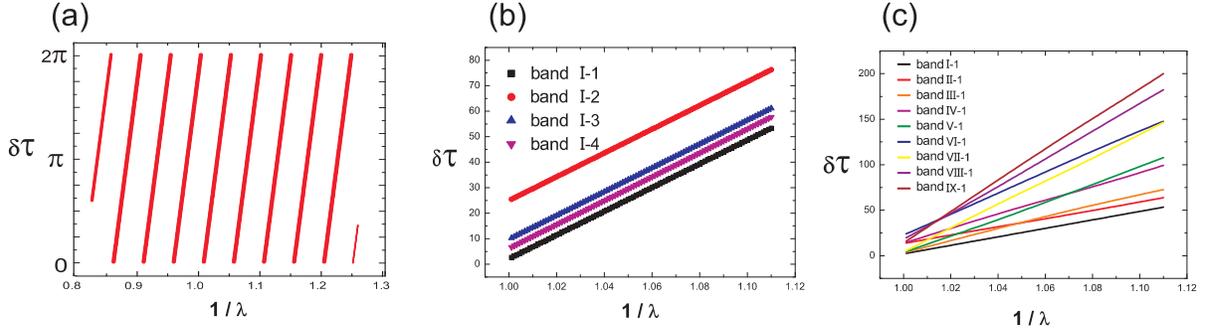}}
\caption{\label{fig AsymmetryFactor}{\bf{Asymmetry factors of
bands.}} a) Asymmetry factor (multiplied by $\tau=2\pi/n$) of the
first band (i.e., the bottom band in the first loop) as function
of $1/\lambda$. b) Extended asymmetry factors (i.e.,
$\delta\tau\in \mathbb{R}$) of the first four bands in the first
loop as function of $1/\lambda$. They all have the same slope. c)
Extended asymmetry factors of the bottom bands in the first nine
loops as function of $1/\lambda$. They have different slopes in
general.}
\end{figure*}

The jumping processes between different branches of solutions give
additional phases to the quantum tunnelling rate $J_l$, which
makes it a complex number $J_l=|J_l|e^{-i\delta_l\tau}$. As we
have discussed in the main text, the connecting condition by
jumping is given by the phase translation operator
$\hat{T}_\tau=e^{-i\tau \hat{a}^\dag \hat{a}}$. Since
$\xi^{(0)}\gg 1$, we can expand operator $\hat{T}_\tau$ by
\cite{Guo} $\hat{a}^\dag
\hat{a}\approx{\xi^{(0)}}/{\lambda}+i\partial/\partial \theta$. As
a result, the connecting condition is
$\hat{T}_\tau\phi_l(\theta)\approx e^{-i{\xi^{(0)}\tau
}/{\lambda}}\phi_l(\theta+\tau)$. Thus we get the symmetry factor
\begin{eqnarray}\label{Jphase}
\delta_l=\delta_l^0+\xi^{(0)} /\lambda,
\end{eqnarray}
where $\delta_l^0$ is the residual asymmetry beyond WKB
calculation. In Fig.~\ref{fig AsymmetryFactor}(a),  we compare the
above linear relationships between $\delta_l$ and $1/\lambda$ for
different bands to our numerical simulations. In Fig.~\ref{fig
AsymmetryFactor}(b) and Fig.~\ref{fig AsymmetryFactor}(c), we
expand the asymmetry factor to the whole field of real number
$\mathbb{R}$ and plot it as function of $1/\lambda$ for different
bands. The bands in Fig.~\ref{fig AsymmetryFactor}(b) are all in
the first loop. We see that, since the states in the first loop
tunnel through the upper boundary, they all have the same slope
given by $\xi^{(0)}$, which is the second zero point of Bessel
function $J_n(\sqrt{2\xi})$. Here, we consider $\xi^{(0)}=0$ is
the first zero point of Bessel function $J_n(\sqrt{2\xi})$ for
$n\neq 0$.

In Fig.~\ref{fig AsymmetryFactor}(c), we show the linear
relationships between $\delta_l$ and $1/\lambda$ for the bottom
bands in different loops. We see their slopes are different. The
reason is that the bands in different loops tunnel though
different paths with different jumping points $\xi^{(0)}$. Like
the states in the first loop, the states in other loops can tunnel
through both the upper boundary and lower boundary. However, we
have checked the integral $\int\mathrm{Im}[\theta]d\xi$ of the
upper path is always larger than that of the lower path.
Therefore, the contribution to the tunnelling rate from the upper
path is exponentially smaller than the contribution from the lower
path. Therefore, the slope of all the bands in the $l$-th ($l>1$)
loop is given by the $l$-th zero point $\xi^{(0)}_l$ of Bessel
function $J_n(\sqrt{2\xi})$. In the flowing table, we compare the
slopes extracted form numerical simulation to our theoretical
calculation.

\begin{center}
     \begin{tabular}{ | l | l | l | l |}
     \hline
     Band Index &  $\xi^{(0)}$ (Numerical) & $\xi^{(0)}$ (Theory) & Relative Errors \\
     \hline
     \Rmnum{1}-1 & 642.241 & 651.545 & -0.014 \\
     \hline
     \Rmnum{2}-1 & 629.514 & 651.545 & -0.034 \\
     \hline
     \Rmnum{3}-1 & 860.600 & 844.308 & 0.019 \\
     \hline
     \Rmnum{4}-1 & 1021.829 & 1032.972 & -0.011 \\
     \hline
     \Rmnum{5}-1 & 1186.088 & 1225.435 & -0.032 \\
     \hline
     \Rmnum{6}-1 & 1427.519 & 1424.378 & 0.002 \\
     \hline
     \Rmnum{7}-1 & 1662.219 & 1631.067 & 0.019 \\
     \hline
     \Rmnum{8}-1 & 1820.811 & 1846.185 & -0.014 \\
     \hline
     \Rmnum{9}-1 & 2056.534 & 2070.142 & -0.007 \\
     \hline
     \end{tabular}
\end{center}

\subsection{Squeezing parameters $v$ and $u$}

In this section, we calculate the squeezing factor $u$ of
localized states near the stable points of phase space lattice.
First, we determine the all the extrema $(r_e,\theta_e)$ including
stable points and unstable saddle points by the derivatives of
quasienergy (\ref{Classicalg}) along both angular direction and
radial direction
\begin{eqnarray}\label{extreme1a}
\frac{\partial g}{\partial \theta}
\big|_{\theta=\theta_e,r=r_e}&=& -2n\mu
J_n(r_e)\sin(n\theta_e-\frac{n\pi}{2})=0,\\ \label{extreme1b}
\frac{\partial g}{\partial r }\big|_{\theta=\theta_e,r=r_e}&=&
\epsilon r_e+\mu
\Big(J_{n-1}(r_e)-J_{n+1}(r_e)\Big)\nonumber\\
&&\ \ \ \ \ \ \ \ \times\cos(n\theta_e-\frac{n\pi}{2}) =0.
\end{eqnarray}
The angular extrema can be obtained from Eq.(\ref{extreme1a}),
that is, $\theta_e=l\tau/2+{n\pi}/{2}$ with $l=0,\pm 1,\pm
2,...,\pm (n-1), n$, where $\tau=2\pi/n$ is defined as
\textit{lattice constant} of phase space lattice. To get the
radial extrema, we need to solve the equation (\ref{extreme1b}).
The stability of these extrema $(r_e,\theta_e)$ is determined by
the second derivatives of $g$. If $({\partial^2 g}/{\partial
\theta}^2)\times({\partial^2 g}/{\partial r}^2)
\big|_{r=r_e,\theta=\theta_e}>0$, the extrema are stable,
otherwise unstable. The second derivatives to angle $\theta$ and
radius $r$ are
\begin{eqnarray}
\frac{\partial^2 g}{\partial \theta^2}
\big|_{\theta=\theta_e,r=r_e}&=&-2n^2\mu
J_n(r_e)\cos(n\theta_e-\frac{n\pi}{2}),\\ \label{derivative1}
\frac{\partial^2 g}{\partial r^2 }\big|_{r=r_e,\theta=\theta_e}&=&
\epsilon+\frac{1}{2}\mu
\Big(J_{n-2}(r_e)+J_{n+2}(r_e)-2J_n(r_e)\Big)\nonumber\\
&&\ \ \ \ \ \
\times\cos(n\theta_e-\frac{n\pi}{2}).\label{derivative2}
\end{eqnarray}
Below, we label the stable points (maxima and minima) and unstable
saddle points by $(r_m,\theta_m)$ and $(r_s,\theta_s)$
respectively. We expand the quasienergy $g$ near the stable points
$(r_m,\theta_m)$ to the second order
\begin{eqnarray}\label{localisedg}
g_{local}&\approx&  g(r_m,\theta_m)+ \frac{1}{2}\frac{\partial^2
g}{r^2_m\partial \theta^2
}\big|_{(r_m,\theta_m)}(r_m\theta-r_m\theta_m)^2\nonumber\\
&&+\frac{1}{2}\frac{\partial^2 g}{\partial r^2
}\big|_{(r_m,\theta_m)}(r-r_m)^2\nonumber\\
&=&g(r_m,\theta_m)+\frac{\tilde{p}^2}{2m_e}
+\frac{1}{2}m_e\omega^2_e\tilde{x}^2.
\end{eqnarray}
Here, we have defined coordinate $\tilde{x}=r-r_m$ and momentum
$\tilde{p}=r_m(\theta-\theta_m)$ near the stable point. The
effective mass $m_e$ and effective frequency $\omega_e$ are given
by $$ m_e=r_m^{2}\Big(\frac{\partial^2 g}{\partial
\theta^2}\Big)^{-1}\big|_{(r_m,\theta_m)}\ \ \mathrm{ and}\ \
\omega_e=\sqrt{m^{-1}_e\frac{\partial^2 g}{\partial
r^2}\big|_{(r_m,\theta_m)}}$$ respectively.

Now, we define the displacement operator
$\hat{D}_\alpha=\exp\Big(\alpha a^\dagger-\alpha^* a\Big)$ and the
squeezing operator $\hat{S}_\xi=\exp\Big[\frac{1}{2}(\xi^* a^2-\xi
a^{\dagger 2})\Big]$, which have the following properties
$$\hat{D}_\alpha^\dagger a \hat{D}_\alpha=a+\alpha, \ \ \ \hat{S}_\xi^\dagger a
\hat{S}_\xi=va+ua^\dagger$$ with $\xi=re^{i\theta}$. The squeezing
parameters are given by $v=\cosh r,\ \ u=-e^{i\theta} \sinh r$ .
We transform the original $\hat{g}$ to localized Hamiltonian
$\hat{g}_{local}$ at the stable point $(r_m,\theta_m)$ by three
operators, i.e.,
$$\hat{g}_{local}=\hat{S}_\xi\hat{D}_\alpha\hat{T}_{\theta_m}\hat{g}\hat{T}^\dagger_{\theta_m}\hat{D}^\dagger_\alpha\hat{S}^\dagger_\xi.$$
Here, we first change the orientation using phase space rotation
operator $\hat{T}_{\theta_m}=e^{-i\theta_m {\hat{a}^\dagger}
\hat{a}}$. Then we move the Hamiltonian to the position of stable
point using displacement operator $\hat{D}_\alpha=e^{\alpha
a^\dagger-\alpha^* a}$. Finally, we squeeze the Hamiltonian to fit
the stable point using squeezing operator $\hat{S}_\xi=e^{[\xi^*
a^2-\xi (a^\dagger)^2]/2}$. By choosing
\begin{eqnarray}
\alpha=-\frac{r_m}{\sqrt{2\lambda}},\ \  \
&&v=\frac{1}{2}\Big(\sqrt{m_e \omega_e}+\frac{1}{\sqrt{m_e
\omega_e}}\Big)\ \  \mathrm{and}\  \nonumber\\
&&u=\frac{1}{2}\Big(\sqrt{m_e \omega_e}-\frac{1}{\sqrt{m_e
\omega_e}}\Big),
\end{eqnarray}
we get the localized Hamiltonian as following
\begin{eqnarray}\label{localisedghighorder}
\hat{g}_{local}&=&\hat{S}_\xi\hat{D}_\alpha\hat{T}_{\theta_m}\hat{g}\hat{T}^\dagger_{\theta_m}\hat{D}^\dagger_\alpha\hat{S}^\dagger_\xi\nonumber\\
&=& \lambda\omega_e(a^\dagger a+\frac{1}{2})+
g(r_m,\theta_m)+o(\lambda^2).
\end{eqnarray}

\subsection{Effective temperature $\bar{n}_e$}

We investigate the quantum dynamics near the bottom of a stable
state. The dissipative dynamics is modified by squeezing and can
be described by an effective temperature $\bar{n}_e$.  The
original master equation is
\begin{equation}\label{ME} \frac{\partial \rho}{\partial
\tau}=-\frac{i}{\lambda}[\hat{g},{\rho}]
 + \kappa(1+\bar{n}){\cal D}[a]{\rho} +  \kappa \bar{n}{\cal
 D}[a^{\dagger}]{\rho}.
\end{equation}
The Lindblad superoperator is defined through ${\cal
D}[A]{\rho}\equiv A{\rho}A^{\dagger}
 -(A^{\dagger}A{\rho}+{\rho}A^{\dagger}A )/2 $, $\bar{n}=(e^{\hbar\omega_0/k_BT}-1)^{-1}$
is the Bose distribution and $\kappa$ is the dimensionless damping
scaled $\omega$. By performing a transformation on the density
operator
$$\tilde{\rho}=\hat{S}_\xi\hat{D}_\alpha\hat{T}_{\theta_m}\rho\hat{T}^\dagger_{\theta_m}\hat{D}^\dagger_\alpha\hat{S}^\dagger_\xi,$$
we transform the master equation (\ref{ME}) into the following
form \cite{GuoPhd}
\begin{eqnarray}
\frac{d\tilde{\rho}}{d\tau}&=&-i[\hat{g}_{local},\tilde{\rho}]
 + \frac{\kappa}{2} \{ (1+\bar{n}_e){\cal D}[a]\tilde{\rho}
    + \bar{n}_e{\cal D}[a^{\dagger}]\tilde{\rho} \}
    \nonumber\\
    &&+\frac{\kappa}{2}M(2a^\dagger\tilde{\rho}a^\dagger-{a^\dagger}^2\tilde{\rho}-\tilde{\rho}{a^\dagger}^2)
    +\frac{\kappa}{2}M^*(2a\tilde{\rho}a-{a}^2\tilde{\rho}-\tilde{\rho}{a}^2).\nonumber\\
\end{eqnarray}
Here, parameter $M=uv^*(2\bar{n}+1)$ is the squeezing number. The
effective Bose distribution is given by
\begin{equation}\label{ne}
\bar{n}_e=\bar{n}|v|^2+(1+\bar{n})|u|^2=|u|^2+\bar{n}(2|u|^2+1).
\end{equation}
Near the bottom of stable points, we can make the harmonic
approximation. The squeezing number $M=uv^*(2\bar{n}+1)$ has no
contribution to the stationary distribution. The ration of
probability over adjoint levels  thus is given approximately by
\cite{GuoPhd} $\bar{n}_e/(1+\bar{n}_e)$.

\end{document}